\documentclass[useAMS,usenatbib]{mn2e}
\usepackage{graphicx}
\usepackage{float}
\usepackage{subcaption}
\usepackage[usenames,dvipsnames]{xcolor}

\title[X-ray time lags in MCG-6-30-15]{The changing X-ray time lag in MCG-6-30-15}
\author[Kara et al.]{E. Kara$^{1}$\thanks{E-mail:
ekara@ast.cam.ac.uk}, A. C. Fabian$^{1}$, A. Marinucci$^{2}$, G. Matt$^{2}$, M. L. Parker$^{1}$, 
\newauthor
W. Alston$^{1}$, L. W. Brenneman$^{3}$, E. M. Cackett$^{4}$ and G. Miniutti$^{5}$ \\
$^{1}$Institute of Astronomy, The University of Cambridge, Madingley Road, Cambridge, CB3 OHA\\
$^{2}$Dipartimento di Matematica e Fisica, Universit\`a degli Studi Roma Tre, via della Vasca Navale 84, 00146 Roma, Italy\\
$^{3}$Harvard-Smithsonian CfA, 60 Garden St. MS-67, Cambridge, MA 02138, USA\\
$^{4}$Department of Physics and Astronomy, Wayne State University, Detroit, MI 48201, USA\\
$^{5}$Centro de Astrobiologia (CSIC-INTA), Dep. de Astrofisica; LAEFF, PO Box 78, E-28691, Villanueva de la Ca{\~n}ada, Madrid, Spain\\
}

\begin{document}

\date{\today}

\pagerange{\pageref{firstpage}--\pageref{lastpage}} \pubyear{2014}

\maketitle

\label{firstpage}

\begin{abstract}
MCG-6-30-15 is one of the most observed Narrow Line Seyfert 1 galaxies in the X-ray band.  In this paper we examine the X-ray time lags in this source using a total of 600~ks in observations (440~ks exposure) taken with the XMM-Newton telescope (300~ks in 2001 and 300~ks in 2013).  Both the old and new observations show the usual hard lag that increases with energy, however, the hard lag turns over to a soft lag at frequencies below $\sim 10^{-4}$~Hz.  The highest frequencies ($\sim 10^{-3}$~Hz) in this source show a clear soft lag, as previously presented for the first 300~ks observation, but no clear iron K lag is detected in either the old or new observation.  The soft lag is more significant in the old observation than the new.  The observations are consistent with a reverberation interpretation, where the soft, reflected emission is delayed with respect to the hard powerlaw component.  These spectral timing results suggest that two distinct variability mechanisms are important in this source: intrinsic coronal variations (which lead to correlated variability in the reprocessed emission), and geometrical changes in the corona.  Variability due to geometrical changes does not result in correlated variability in the reflection, and therefore inhibits the clear detection of an iron K lag.  
%  These spectral timing results suggest that MCG-6-30-15 has a compact, yet extended corona that changes in time, and that gravitational light bending effects cause the high-frequency variability to be smeared out as it reflects off the inner accretion disc.   
\end{abstract}


\section{Introduction}
\label{intro}

The well-studied Narrow Line Seyfert I galaxy MCG-6-30-15 was the first source where a broad iron K emission line was found \citep{tanaka95}. Since the initial discovery with ASCA, the source has been observed several times with a number of instruments, and the broad spectral feature has been well studied \citep{fabian03, brenneman06, marinucci14}.  It also shows clear evidence for warm absorbers through complex absorption structures \citep{otani96,lee01,chiang11}. In addition to these striking spectral features, the source shows significant X-ray variability on a range of timescales \citep{vaughan04, papadakis05, emm11, parker14}.  In this paper, we examine the X-ray time lags from two 300~ks observations with XMM-Newton, from 2001 and 2013.

The X-ray time lags in this source were originally studied using the first 300~ks observation by \citep{emm11}, who found a significant `soft' lag at high frequencies and a `hard' lag at low frequencies.  This was the second source found with such a significant soft lag, the first of which was 1H0707-495 \citep{fabian09}. In that work, the soft lag was interpreted as the light travel reverberation time between the primary X-ray continuum emission and the reprocessed emission off the inner accretion disc at a few gravitational radii from the central black hole.  \citet{emm11} fitted the lag of MCG-6-30-15 with impulse response functions, and also found that the reverberation interpretation could well describe the data.  Since these initial discoveries, soft lags have been found in about 20 sources \citep{demarco11,zoghbi11,cackett13}, and the frequency and the amplitude of the lag has been found to scale with the black hole mass, indicating that the light travel time between the X-ray corona and the disc is small, i.e. less than $\sim$ 10 gravitational radii \citep{demarco13}.

\begin{figure*}
\begin{subfigure}{\columnwidth}
\includegraphics[width=\columnwidth]{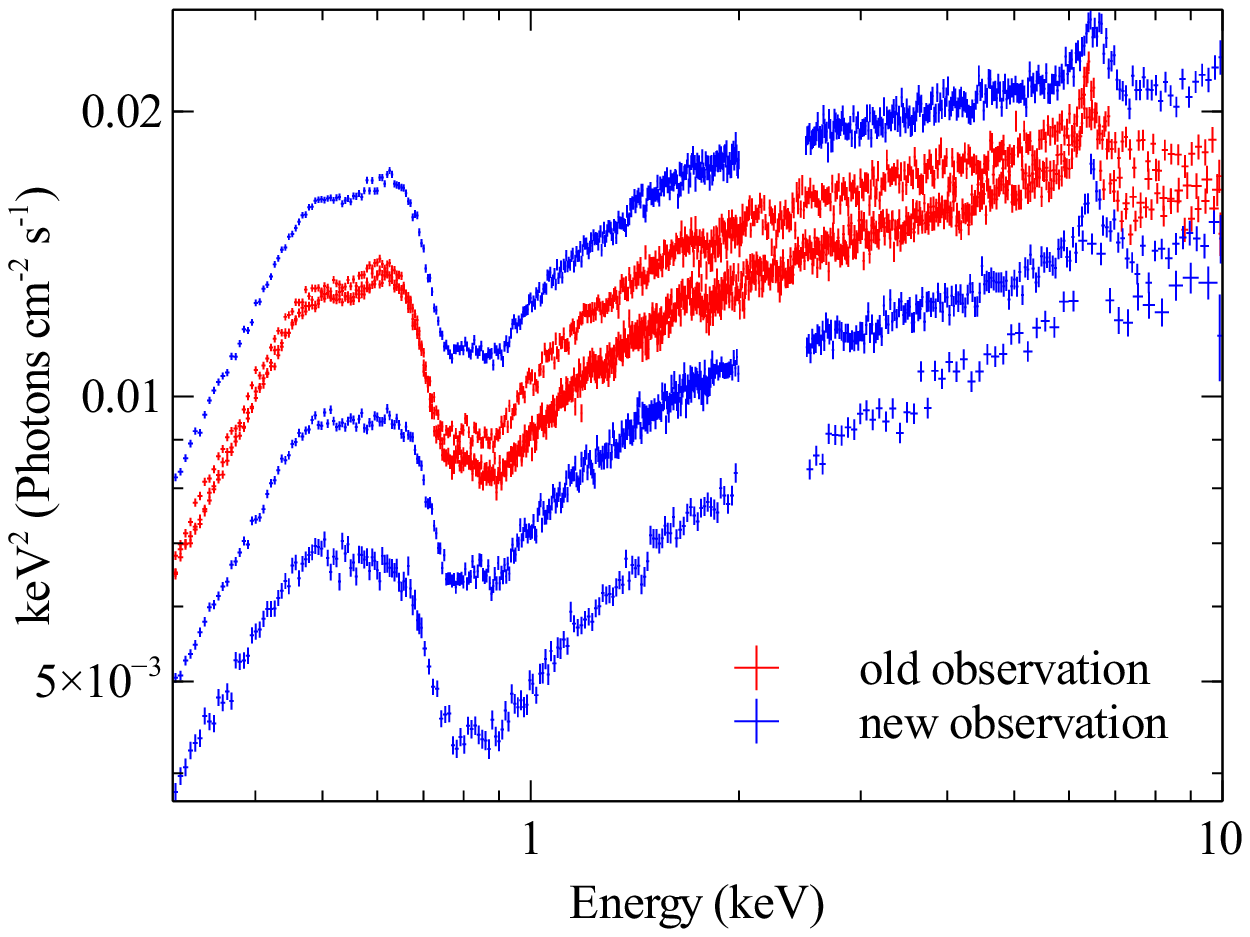}\\
\end{subfigure}
\begin{subfigure}{\columnwidth}
\includegraphics[width=\columnwidth]{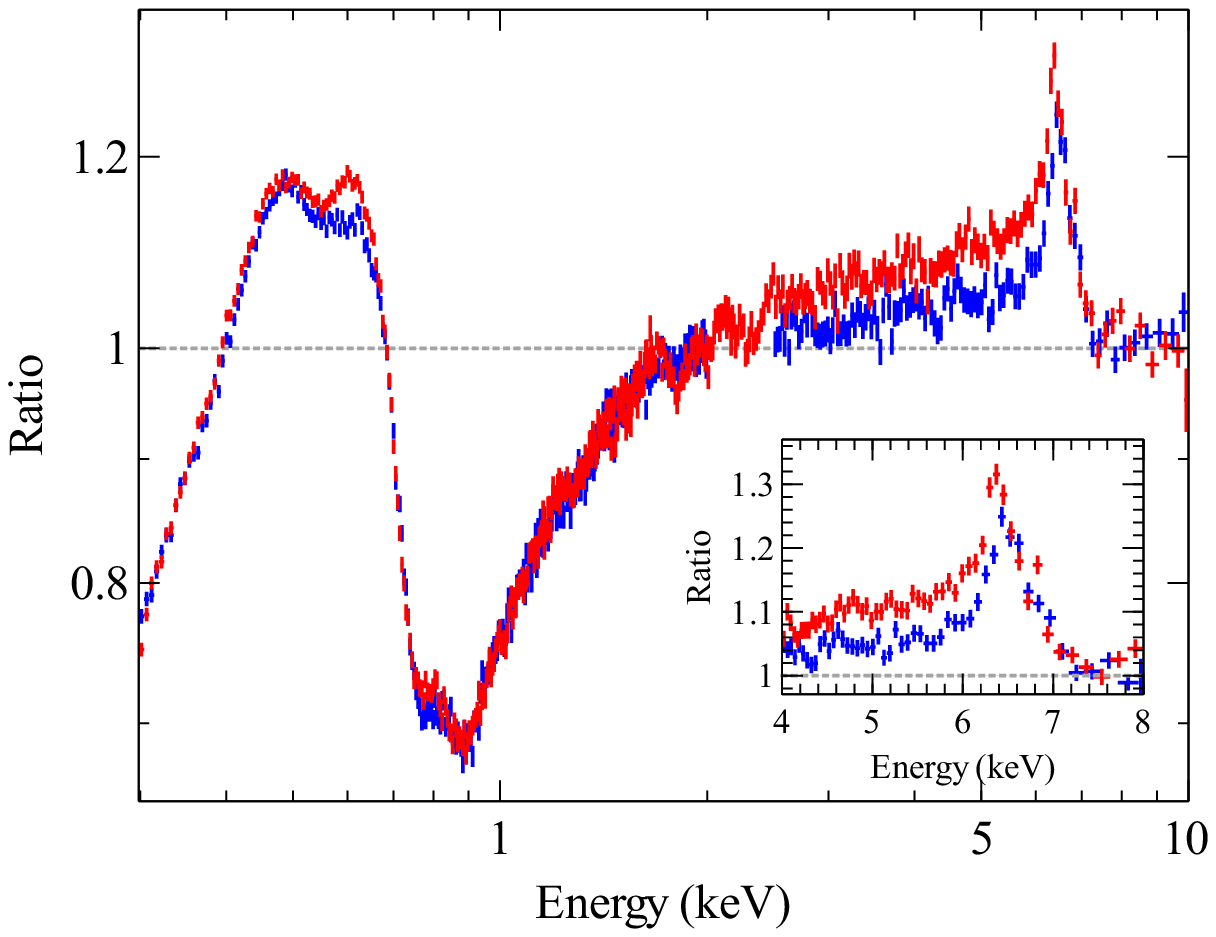}\\
\end{subfigure}
\caption{({\em Left:}) The mean spectrum for individual orbits for the old observation in red and the new in blue. We maintain this color scheme for the old and new observations throughout the paper.  ({\em Right: }) The ratio of the mean spectrum to a powerlaw model fit where the continuum dominates from 1.9--2~keV and 7.5--10~keV.  This avoids the warm absorption at soft energies, a calibration discrepancy at 2--2.5~keV, and the broad Fe~K emission line peaking at 6.4~keV.  It is clear from this ratio plot that the red wing of the Fe~K line is stronger in the old observation than in the new.  The plot inset shows a zoom in of the ratio from 4--8~keV.  From this we see that the red wing of the Fe~K line is stronger in the old observation, but the blue wing (largely dependent on the disc inclination) is the same.}
\label{spec}
\end{figure*}

The strongest evidence for reverberation came with the discovery of the Fe~K lag. This was first found in NGC~4151 \citep{zoghbi12}, where the light travel time between the continuum and the gravitationally redshifted wing of the line is shorter than the time between the continuum and the line centroid, produced at larger radii.  Fe~K lags have been discovered in 10 NLS1 sources thus far \citep{kara13a,kara13b,zoghbi13,kara13c,marinucci14}. The soft lag and Fe~K lag have been modelled using general relativistic ray tracing simulations, and are also consistent with a compact corona illuminating an ionised accretion disc that extends to the innermost stable circular orbit \citep{cackett14,emm14}.  The natural step is to look for an Fe~K lag in MCG-6-30-15, the original source with a broad Fe~K line.

In addition to the high-frequency reverberation lags that show a clear signature of relativistic reflection, there is a very different process at low frequencies.  At low frequencies, the hard emission is seen to lag behind the soft by several times longer than the reverberation delay.  The lag-energy spectrum does not show a reflection feature, rather it has a log-linear increase with energy, and shows no spectral features \citep[e.g. the lags in Ark~564; ][]{kara13c}. These lags have been observed in black hole binaries (BHB) for decades \citep{miyamoto89, nowak99}, and have also been observed in AGN \citep{papadakis01,mchardy07, arevalo06b}.  The interpretation of the low-frequency lag is not well understood,  but the leading interpretation is that mass accretion rate fluctuations in the accretion disc propagate inwards on the viscous timescales \citep{kotov01}. This is a multiplicative effect, which causes the correlated variability at a range of radii.  These disc fluctuations cause low-frequency variability that is transferred up to the corona, likely via magnetic field lines.  If the corona is slightly extended (i.e. not a point source) and the soft emission originates from larger radii than the hard, then the disc fluctuations will cause the soft emission to respond before the hard, thus causing the observed `hard' lag \citep{arevalo06}. 

In this paper, we examine the high and low-frequency lags using 600~ks of XMM-Newton observations.  The paper is constructed as follows: in Section~\ref{obs} we present the data reduction and describe the two observations that we are working with (one 300~ks observation from 2001, and one 300~ks observation taken in 2013).  In Section~\ref{results}, we show the results of our variability analysis, by examining how the low and high frequency variability and how it changes between observations. We do this through an analysis of the coherence, the covariance and the time lags. We conclude with a discussion of the results in Section~\ref{discuss}.

\begin{figure*}
\includegraphics[width=\textwidth]{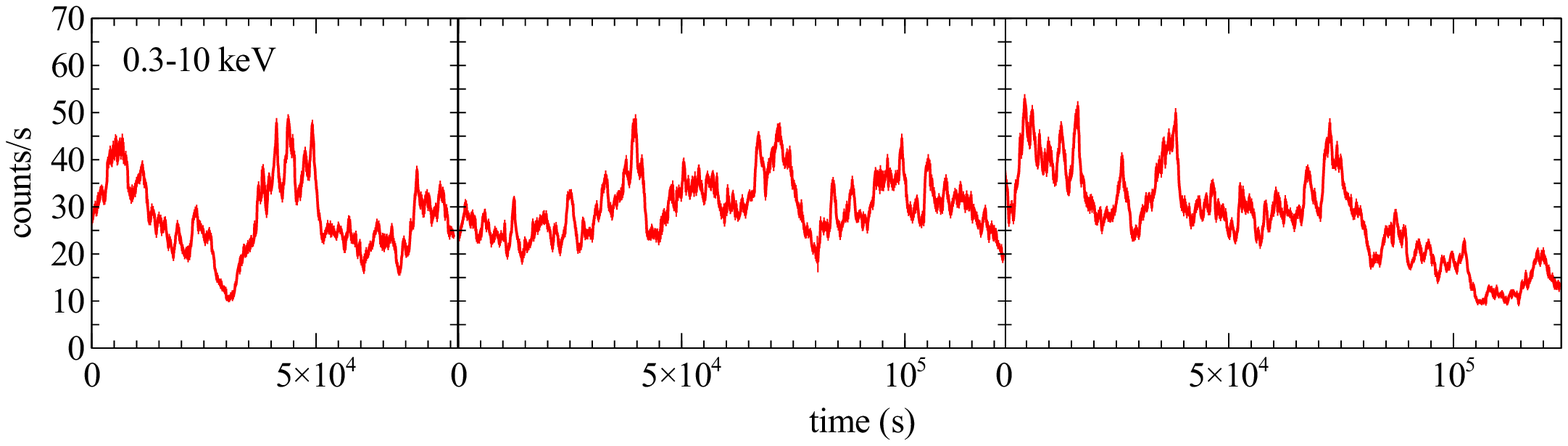}
\includegraphics[width=\textwidth]{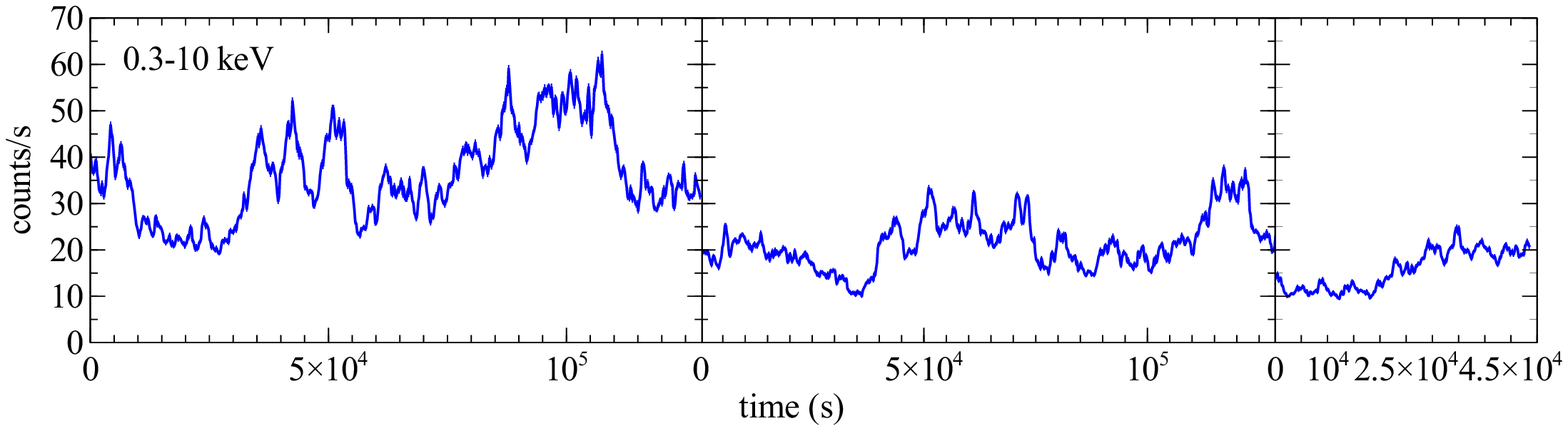}
\caption{The 0.3--10~keV light curves for the old observations taken in 2001 (top) and the new observations from 2013 (bottom).  Both show rapid variability, and are at roughly the same flux level.  We will present the lag analysis for these two observations separately, as there appears to be some non-stationarity between the observations.}
\label{lc}
\end{figure*}

\section{Observations and Data Analysis}
\label{obs}

MCG-6-30-15 was observed with the XMM-Newton telescope in 2001 and, more recently in 2013.  In this work, we analyze both data sets in the same way with the newest calibration files. We focus on the data from the EPIC-PN camera.  

The {\em XMM-Newton} satellite \citep{jansen01} observed MCG-6-30-15 for 300~ks over three orbits from 2001 July 31 to 2001 August 4 (Obs. IDs 0029740101, 0029740701, 0029740801) and again for 300~ks over three orbits from 2013 January 29 to 2013 February 2 (Obs. IDs 0693781201, 0693781301, 0693781401). For this timing analysis, we focus on the high time-resolution data from the EPIC-pn camera \citep{struder01}. The observations were taken in small window imaging mode.

The data were cleaned for high background flares, which resulted in a total exposure time of 228~ks and 215~ks for the old and new observations, respectively.  The data were selected with the condition {\sc pattern} $\le 4$. Pile-up effects were not significant in any of the observations.
                                                                                                                                                  
The source light curve was extracted from circular regions of radius 35 arcsec, which were centered on the maximum source emission. The background light curves were chosen from a circular  region of the same size, and were the same distance to the readout node as the source region.  The background subtracted light curves were produced using the tool {\sc epiclccorr}. 
                                                         
The left panel of Fig.~\ref{spec} shows mean counts spectra for all three observations from 2001 in red and from 2013 in blue. We ignored data between 2-2.5 keV due to known calibration effects (Smith et al. 2013\footnote{available at {\tt http://xmm2.esac.esa.int/docs/documents/\\CAL-SRN-0300-1-0.pdf}}). The spectrum for the old observation remained fairly constant between the three orbits, while the new observation shows more variability between the orbits. We take a closer look in the right panel of Fig.~\ref{spec} where the total mean spectra from the old and new observations are fit with power laws in the 1.9--2~keV and 7.5--10~keV band \citep[following the same procedure done for the new observations in ][]{parker14}.  These energy bands allow us to avoid the warm absorber at soft energies, the calibration effect at 2--2.5~keV, and the broad Fe~K emission line.   Below $\sim 1.5$~keV, the two spectra look similar, but above 3~keV, we see that the red wing of the Fe~K line is much stronger in the old observation than in the new. We zoom into Fe~K band in the inset.  From this we see that the peak of the Fe~K line is at slightly lower energies in the old observation, but that the blue wing of the Fe~K line remains the same between the two. We note that even looking at the line profiles from individual orbits shows that all of the new orbits (including the highest flux orbit) have narrower lines than the old observations. This ratio plot is our first hint of the difference between the old and observations.

\label{results}
\begin{figure*}
\includegraphics[width=\textwidth]{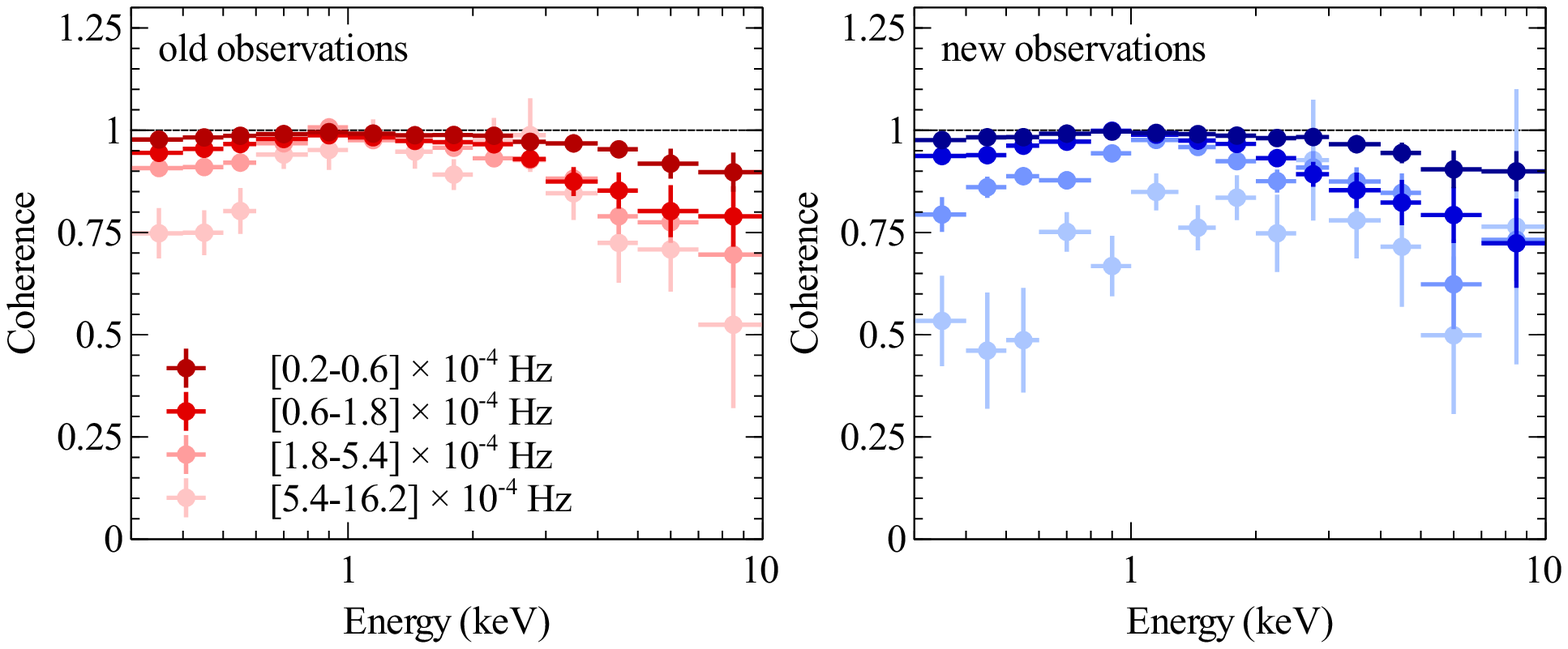}
\caption{The coherence spectra at different frequencies for old observations (left) and new observations (right). The colors increasing from dark to light show the spectra for increasing frequency.   The maximum coherence is one, and the higher the coherence, the more reliable the lag measurement. In the old observations the coherence is high at all frequencies. The coherence is high in the new observations as well, but the highest frequency clearly has a lower coherence than in the older observations.}
\label{coherence}
\end{figure*}

Fig.~\ref{lc} shows the 0.3--10~keV light curves for the old observations (top; red) and new (bottom; blue). The two observations look similar, both showing a high flux and rapid variability.  The new observation has greater spectral variability. As the flux decreases, so does the variance, as is often seen in AGN and BHB \citep{uttley05}.

These are the observations that we will analyse in this paper. First we compare the variability of the two observations by looking at the coherence spectra and covariance spectra. Then we finish by completing a lag analysis.

\section{Results}

\subsection{Coherence}

We use the coherence as a check for  whether a reliable measurement of the lag can be taken at a particular frequency.  The coherence calculates to what degree one light curve is a simple linear transformation of the other \citep{vaughan97}.  A coherence of 1 indicates that they are complete linear transforms of each other.  The coherence must be high (though not necessarily 1) in order to reliably measure the lag \citep{kara13a}.   Fig.~\ref{coherence} shows the coherence at several frequency ranges for the old observations on the left (in red) and the new observations on the right (in blue).  The lowest frequencies are darkest, and the frequency increases as the color lightens.  The coherence in this case is measured between the light curve in each small energy bin, and a broad reference light curve, which we choose to be the entire 0.3--10~keV band light curve, with the energy bin of interest removed, so that noise is not correlated.  We find that the coherence is high at very low frequencies in both observations. The coherence looks similar up until the highest frequency band, where the new observation has a significantly lower coherence than the old observation.  While the coherence does not need to be 1 in order to measure a lag, it does indicate that either the variability is lower in the new observation, and we are closer to the level of the Poisson noise, or that there is some additional nonlinear complexity occurring at higher frequencies.  This will be important later when looking at the high-frequency lags.

\subsection{Frequency-resolved covariance spectra}

\begin{figure*}
\includegraphics[width=\textwidth]{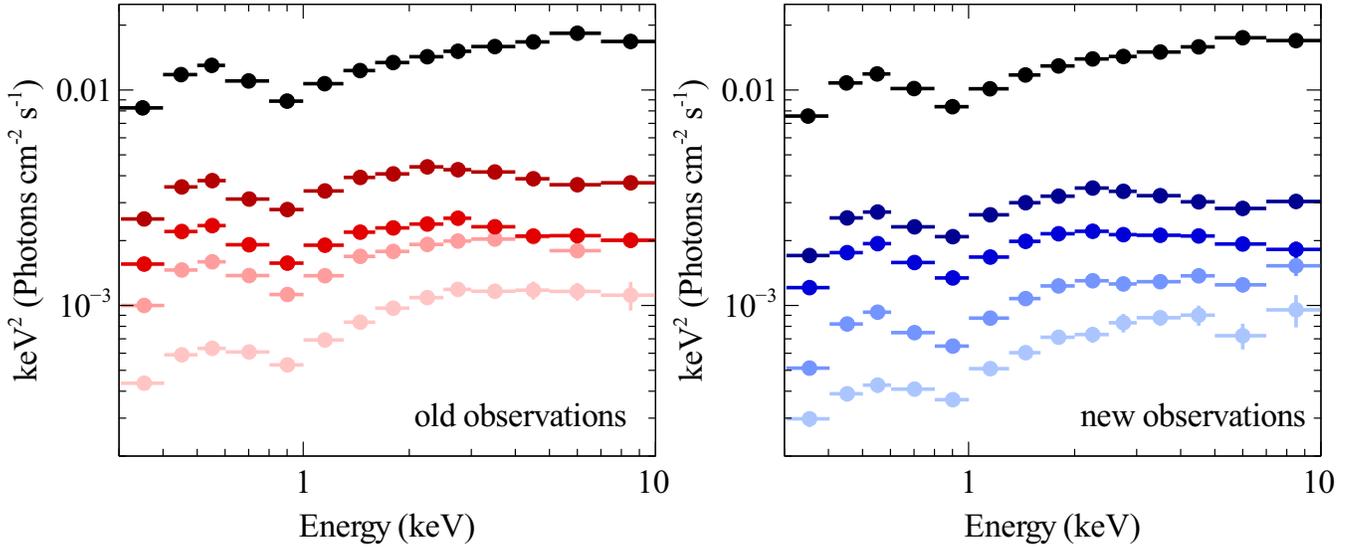}
\caption{The frequency-resolved energy spectra (a.k.a. the covariance spectrum) for the old observations (left) and the new observations (right). As in Fig.~\ref{coherence}, the frequency increases from dark to light colors, as indicated in the key. The black spectra shows the time integrated energy spectrum for comparison.  The spectrum looks similar at all frequencies, but it is clear that there is less variability at all frequencies in the new observation than the old.  Also evident from the plots is that there is an excess of emission in the mean spectrum that is not present at any of the frequency resolved spectra, indicating that the emission around the broad iron line is not varying at all frequencies probed.}
\label{covariance}
\end{figure*}

\begin{figure*}
\includegraphics[width=\textwidth]{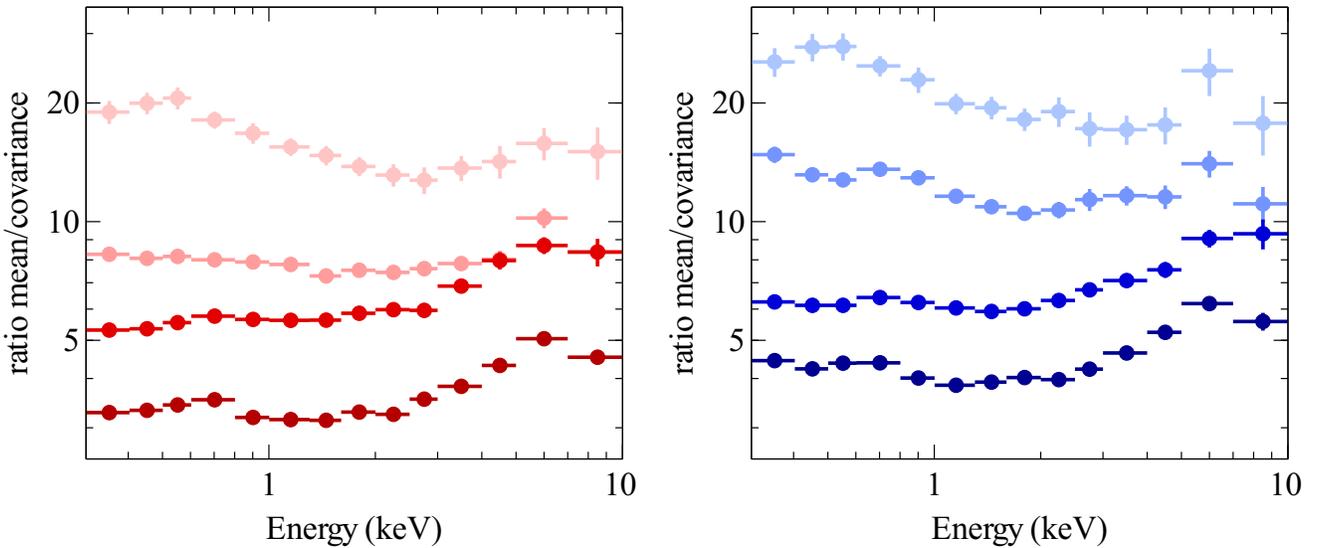}
\caption{The mean spectrum divided by the covariance spectra at the different frequencies, as shown in the previous figure.  This is essentially showing the spectrum that does {\em not} vary at each frequency range.  Each frequency range shows that the Fe~K line does not vary at that particular frequency.  There is clearly also a soft excess that is not varying at each frequency.}
\label{cov_rat}
\end{figure*}

Next we examine the frequency-resolved energy spectra (or covariance spectrum) to see what part of the variable spectrum is changing.  We compute the covariance spectrum in the frequency domain, as outlined in \citet{uttley11}.  The covariance spectrum is similar to the rms spectrum, except that it picks out only the {\em correlated} variability (which make it easier to compare with the lag that is also calculated just between the correlated variability).  The covariance spectrum is a measure of the absolute amplitude of correlated variations in count rate as a function of energy \citep{wilkinson09}. Furthermore, as the covariance is measured in count rate, we can more easily compare it to the mean energy spectrum.

We compute the covariance with respect to the full 0.3--10~keV reference band (as with the coherence). Fig.~\ref{covariance} shows the covariance for the same frequency ranges shown in Fig.~\ref{coherence}.  The black spectrum on the top shows the mean spectrum coarsely binned to the same binning as the covariance spectra. The covariance spectra shown increase in frequency from dark to light.  For the covariance spectra and the mean spectrum we show the unfolded spectra to a powerlaw model with index 0 and normalisation 1.

The first thing to notice is that the new observations on the right have lower correlated variability at all frequencies probed, and especially at the highest frequency range. Even though the new observation shows more variability between orbits, on the timescale of individual orbits and shorter, the old observation has more correlated variability.  

We also notice that while the mean spectrum increases steadily above 1~keV, in the variable spectrum the peak of the emission is around 2~keV, beyond which it flattens off, or even decreases.  This is showing that while there is a clear broad iron line in the mean spectrum, it is not apparent in the variable spectrum at any frequency.  This effect has been seen at low frequencies in MCG-6-30-15 \citep{vaughan04, papadakis05}, and now we show that the same occurs at high frequencies in this source.  
 
This effect is seen more clearly by taking the ratio of the mean spectrum to the covariance spectra at different frequencies, as shown in Fig.~\ref{cov_rat}.  These mean-to-covariance ratio plots show the part of the energy spectrum that does {\em not} show correlated variability at a particular frequency.  The shape of these ratio plots for each frequency range resembles a broad iron line profile and a soft excess.  Again, this indicates that the iron line (and possibly the soft excess) contribute to the mean spectrum, but not much to the covariance spectrum at the frequencies probed.  

\subsection{Time lags}
\subsubsection{Low-Frequency lags}

\begin{figure}
\includegraphics[width=\columnwidth]{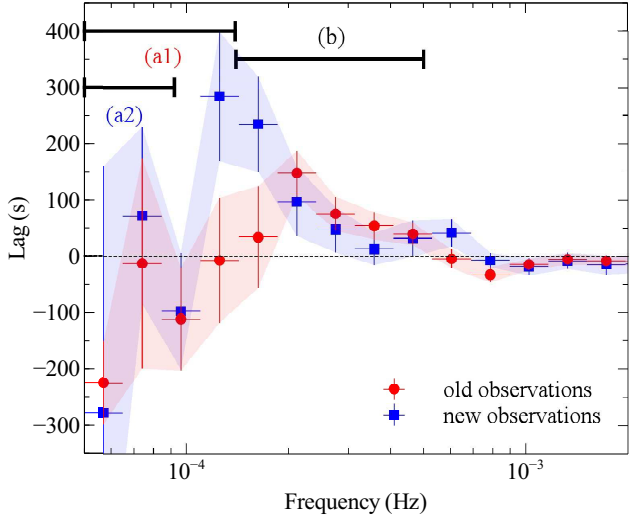}
\caption{The lag-frequency spectrum between 0.3--1.6~keV and 1.6--5~keV for old and new observations, in red and blue, respectively.  Above $\nu>2 \times 10^{-4}$~Hz, we see a hard lag, as is found in most NLS1 sources and X-ray binaries.  Below this frequency, the lag sharply disappears in both observations. The frequency of the downturn is slightly higher in the old observation.  We explore frequencies denoted (a) and (b) by looking at their lag-energy spectra.}
\label{low_lagfreq}
\end{figure}

\begin{figure}
\includegraphics[width=\columnwidth]{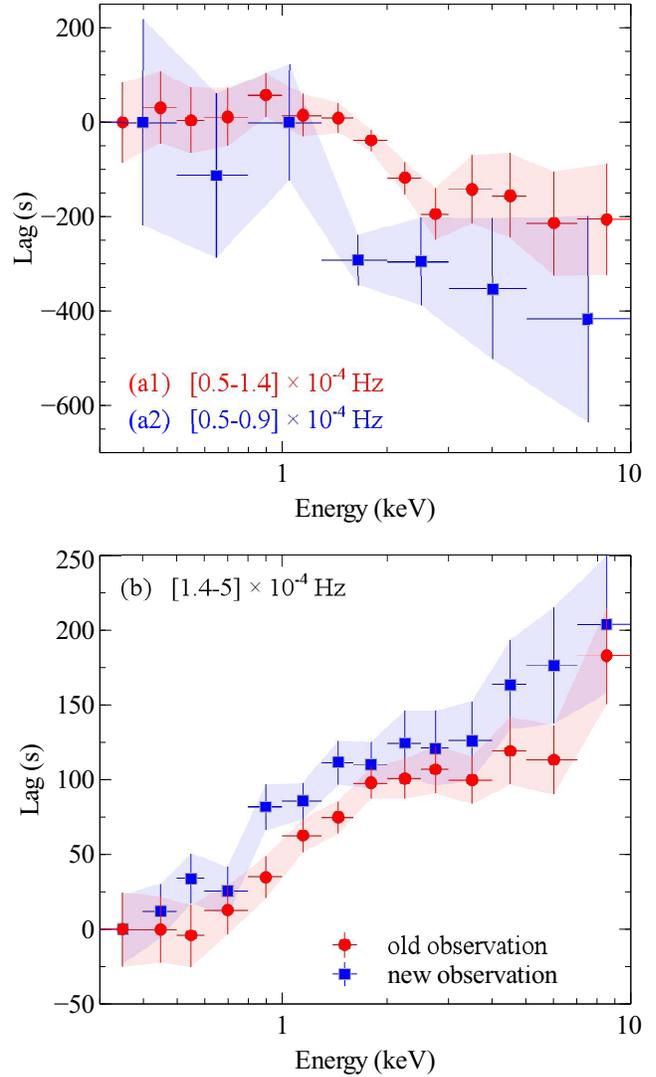}
\caption{The low-frequency lag-energy spectra for old observations (in red) and new observations (in blue) for very-low frequencies ($\nu=[0.5-1.4] \times 10^{-4}$~Hz, for the old observations and $\nu=[0.5-0.9] \times 10^{-4}$~Hz for the new) on the top, and low-frequencies ($\nu=[1.4-5] \times 10^{-4}$~Hz for both observations) on the bottom.  The low frequency on the bottom looks like the normal featureless increase with energy that is found in many NLS1s and X-ray binaries.  It is similar for the old and new observations.  The very-low frequencies on the top show clear soft lags, that are similar in both amplitude and shape, though the new observations have larger error bars (due to the narrower frequency range probed).}
\label{low_lagen}
\end{figure}

With an insight into the variability of MCG-6-30-15 in these two observations, we now examine the time lags on long and short timescales.

We compute the lags with the standard Fourier technique outlined in \citet{nowak99}.  To describe briefly, we prepare light curves in particular energy bins with 10~s time bins.  We do not break the light curves into smaller segments as we want to probe the longest timescales possible.  We compute the Discrete Fourier Transform of each light curve, and compute the cross spectrum between two light curves. The cross spectrum is defined as the DFT of one light curve times the complex conjugate of the DFT of the other. It therefore contains a real amplitude and a complex exponential phase (which is the phase difference between the two light curves). This cross spectrum is then averaged over a frequency band, with midpoint $f$. We divide the phase lag by $2 \pi f$ to obtain a frequency-dependent time lag for each frequency bin.  The lag-frequency spectra shown are the frequency dependent lags between two large energy bands \citep[in this case between the soft band 0.3--1.6~keV and the hard band, 1.6--5~keV, as in ][]{emm11}.  The lag is defined such that positive means that the hard band lags the soft.  The lag-energy spectra look at the energy dependence of the lag at a broad frequency band (which we choose based on the lag-frequency spectrum).  The lag is computed for each channel of interest with respect to a large reference band. For this work, we choose the reference band to be the entire 0.3--10~keV band in order to maximize the signal-to-noise.  We remove each channel of interest from the reference band so that noise is not correlated. The lag is read from bottom to top, i.e. the smaller the lag the earlier the emission arrives at our detector.  We are interested in the relative lag difference between energy bins, and not in the absolute amplitude of the lag, as that is simply determined by the choice of reference band.

Fig.~\ref{low_lagfreq} shows the lag-frequency spectra between the soft (0.3--1.6 keV) and hard (1.6--5~keV) light curves for the old observation in red and the new observation in blue. (If we choose the `typical' soft band from 0.3--1 keV and the hard band from 1--4~keV, the lag results are much the same, but the signal is lower, as the energy bands are smaller.)  These lag-frequency spectra focus on the low-frequency lags, and we will zoom in to the high frequency lags later.  First looking at the old observation  above $\nu>2 \times 10^{-4}$~Hz, as presented by \citet{emm11}, we find the typical hard lag of about 200~s at its maximum.  This behavior is seen often in NLS1 sources and in BHB \citep{mchardy07,papadakis01,miyamoto89}.  Below this frequency, however, we see a break, and the lag disappears, or even becomes negative, indicating that the soft band leads the hard band at the lowest frequencies.  The lowest frequencies correspond to a timescale of 5-20~ks. 

This is not an effect of the finite size of the observation, as the turnover occurs at a sufficiently high frequency to make a reliable measurement of the lag.  We confirm this through simulations. We compute Monte Carlo \citet{timmer95} light curves from the observed Power Spectral Density (PSD) index, scaled them to the count rate, variance and exposure of our observations, and added Poisson noise.  We simulated 1000 light curve pairs offset by a constant phase, and measured a reliable lag down to $5 \times 10^{-5}$~Hz. This confirms that the negative lag seen at low-frequencies in all the observations is real.

This behavior has only been seen in two Seyfert galaxies, NGC~3516 \citep{turner11} and NGC~4051 in the low-flux state \citep{alston13}, though neither of those cases showed a hard lag at low-frequencies and a soft lag at lower frequencies in the same observation, as we see in MCG-6-30-15. We will discuss the possible origin of this very-low-frequency soft lag in Section~\ref{discuss}.

\begin{figure}
\includegraphics[width=\columnwidth]{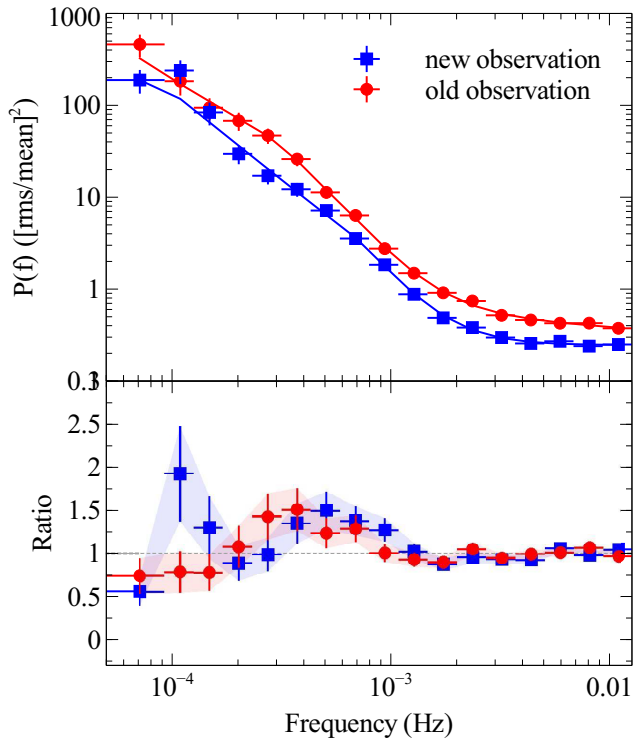}
\caption{The 0.3--10~keV power spectral densities for the old and new observations.  The dotted lines show the best fit twice broken powerlaw models to the data.  we fit a high-frequency break at $\sim 7 \times 10^{-4}$~Hz in both observations, and a low-frequency break at $3 \times 10^{-4}$~Hz for the old observation and $9 \times 10^{-5}$~Hz for the new observation.  The low-frequency breaks occur at similar frequencies as the turnover frequency for the lags. The lower panel highlights the structure in the PSD by showing the ratio of the data to a single powerlaw model (with additional white noise).}
\label{powspec}
\end{figure}

The new observation in blue shows similar behavior of the usual hard lag that disappears at the lowest frequencies. Interestingly, the break frequency is lower in this observation, and it is not clear whether the lag becomes negative at the lowest frequencies or if it just goes to zero.

To examine the low-frequency hard and soft lags further, we construct the lag-energy spectrum for the frequency ranges (a1), (a2) and (b) indicated in Fig.~\ref{low_lagfreq}.  We probe slightly different very-low-frequency lags, (a1) and (a2), for the old and new observations, because the break in the lag occurs at slightly different frequencies.  The corresponding lag-energy spectra are shown in Fig.~\ref{low_lagen}.  The panel on the top shows the very-low-frequency soft lags for the old observations in red and the new in blue.  Both observations show a clear soft lag.  For the old observations, there is zero lag between the emission from 0.3--2~keV, but above 2~keV, the lag drops by 200~s.  The lag appears to plateau at above 3~keV, though it is difficult to see with such large error bars.  The behavior in the new observation is similar, though the dip appears at slightly lower energies, around 1~keV.  Combining the old and new observations does not improve the signal, as the new observations have a lower signal-to-noise, and just act to smear out the clearer signal from the old observation.  In total, we have 600~ks in 6 orbits. Each individual orbit shows this very-low-frequency soft lag, which suggests that this is not a transient phenomenon.  

At slightly higher frequencies (b), in the bottom panel of Fig.~\ref{low_lagen}, we are probing the hard lag.  The lag looks very similar between old and new observations, and also looks very similar to the hard lags found in earlier lag studies.  The lag-energy spectrum shows a log-linear increase with energy. 

\subsubsection{Power Spectra}

For completeness, we check the PSD of both the old and new observations.  \citet{vaughan03}, using the old 300~ks {\em XMM-Newton} observation, found that while the data are sparse, the PSD can be well fit with a break at around $10^{-4}$~Hz, similar to the frequency we found for the break in the lags.  This break frequency was consistent with other PSD analyses of MCG-6-30-15 using RXTE data \citep{uttley02, mchardy05}. \citet{vaughan03} also found that the PSD above the break frequency is energy dependent, with the hard band exhibiting a harder PSD (i.e. more power at higher frequencies).  Below the break frequency, they fixed the slope to be -1 for all energy bands.  \citep{mchardy05} was able to improve constraints on the low-frequency PSD using long RXTE observations, and found that the low-frequency PSD slope was slightly flatter, but did not find evidence for it changing with energy.

We estimate the PSD using standard methods \citep{vanderklis89,vaughan03} using an rms normalisation.  The top panel of Fig.~\ref{powspec} shows the broadband PSD from 0.3--10~keV for both the old and new observations.  The frequency binning shows equally spaced logarithmic bins, except at the lowest frequency bin, where we have combined two frequency bins so that there were at least 10 frequency samples in the bin.  Qualitatively, we see that the overall normalisation of the new observation is less at all frequencies (as evident from the covariance spectrum).  At high frequencies we are dominated by white noise. The slope is similar between the two, but there are some differences in the shape.  The bottom panel of Fig.~\ref{powspec} shows the ratio of the data to a simple power law model (with a power law of index 0 to describe the white noise).  We see from the ratio plot that the new observation has a clear peak starting at around 2--4 $\times 10^{-4}$~Hz, and possibly a second peak at around 7 $\times 10^{-4}$~Hz.  The new observation shows a clear peak at $10^{-4}$~Hz, and a second, broader peak at around 4--8 $\times 10^{-4}$~Hz.  From this ratio plot, if we assume a twice broken powerlaw model for both the old and new observations, we fit a high-frequency break at $\sim 7 \times 10^{-4}$~Hz in both observations, and a low-frequency break at $3 \times 10^{-4}$~Hz for the old observation and $9 \times 10^{-5}$~Hz for the new observation.  The low-frequency breaks are different between observations, and the break in the new observation occurs at a lower frequency, just as the turnover of the lag. Unfortunately, due to sparse data, we cannot use finer frequency bins than this to better constrain the low frequency break, especially for the old observation.  More sophisticated modelling of the PSD is beyond the scope of this work, though we can see from the ratio plot that the shape of the PSD changes between observations.

\subsubsection{High-Frequency Lags}

\begin{figure}
\includegraphics[width=\columnwidth]{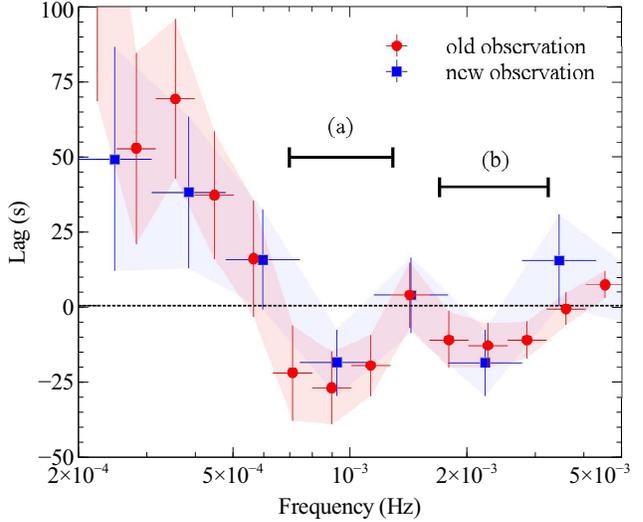}
\caption{The lag-frequency spectra between 0.3--1.6~keV and 1.6--5~keV (same as in Fig.~\ref{low_lagfreq}), now probing higher frequencies.  The old observations are shown in red, and the new are shown in blue.  The old observations (with more high-frequency variability and higher coherence), shows a much `cleaner' lag than the new observations, where a soft lag is tenuous.  The old observations show a clear double trough structure. We will explore these two frequency ranges, denoted (a) and (b), through their lag-energy spectra.}
\label{high_lagfreq}
\end{figure}

\begin{figure}
\includegraphics[width=\columnwidth]{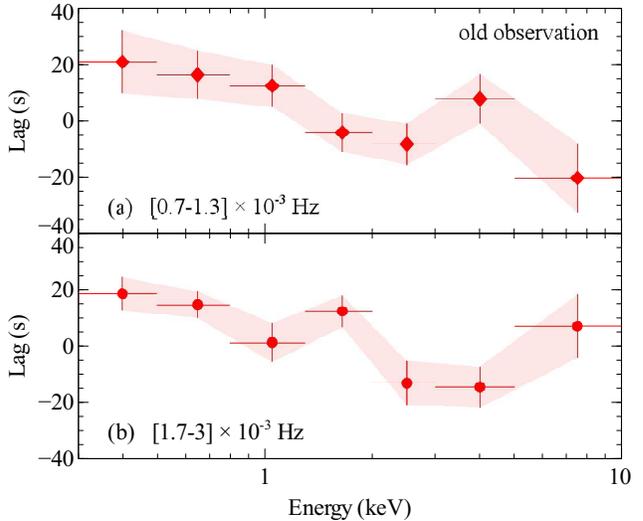}
\caption{The high-frequency lag-energy spectrum for frequency range (a) from $[0.7-1.3] \times 10^{-3}$~Hz on the top, and for slightly higher frequency range (b) from $[1.7-3] \times 10^{-3}$~Hz on the bottom.  At such high frequencies we are close to the level of the noise, and so we cannot put strong constraints on the shape of the lag-energy spectrum, but the overall shape is similar to what we see in NLS1s. There is no clear Fe~K lag, though there is an increase in both frequency bands above 3~keV }
\label{high_lagen}
\end{figure}

Next we examine the lag-frequency spectrum at high frequencies (Fig.~\ref{high_lagfreq}).  The energy bands are the same as in Fig.~\ref{low_lagfreq}, but now we are just zooming in to the higher frequency lags, and are using finer frequency bins.  The red points shows the old observations, and the blue points shows the new observations.  The old and new observations looked similar at low frequencies (in their lags, but also in the covariance and coherence), but the observations look very different at high frequencies.  The lag is much clearer in the old observation, and is barely significant in the new. In the new observation, the variability is lower at this frequency, and so we are close to the level of the noise. It was not possible to delve further by looking at the lag-energy spectrum in any detail.  

We can, however, look closer at the lags in the old observation.  The red points in Fig.~\ref{high_lagfreq} shows a clear double trough structure to the lag.  This is consistent with earlier high-frequency lag studies presented by \citet{emm11} and later by \citet{demarco13}.  We note that the double trough in the lag is not coincident with dips in the PSD at these frequencies (see the PSD in Fig.~\ref{powspec}).  This suggests that the double trough cannot be due to a ringing effect from a single transfer function, as this would be accompanied by dips in the PSD \citep[See Fig.~22 comparing the lags and the PSD for a single transfer function in][]{uttley14}. 

 For completeness, we show the lag-energy spectrum of each trough in the frequency ranges indicated as (a) and (b) in Fig.~\ref{high_lagfreq}.  Fig.~\ref{high_lagen} shows the lag-energy spectra for frequency range (a) on the top and frequency range (b) on the bottom. We cannot use finer energy bins than this due to low statistics. We find that the shape of the high-frequency lag-energy spectra of the old observations are consistent between the two troughs, and look similar to what we find in other high-frequency lag-energy spectra \citep[e.g. 1H0707-495 in ][]{kara13a}. We notice, however, that there is no clear Fe~K lag in these observations. While there is a hint of an upturn in both frequency ranges, there is no clear Fe~K lag.  Even when we combine the old and new observations (for a total of 600~ks of data), or when we examine just the highest flux observations (i.e. the old observations plus the first orbit of the new observations), we do not find a Fe~K lag. We discuss possible reasons for this in Section~\ref{discuss}.

\section{Discussion}
\label{discuss}

We have presented the low and high frequency variability in the old and new observations of MCG-6-30-15. In particular we find:

\begin{enumerate}
\item The mean spectrum shows that the old observation has stronger Fe~K emission line (right panel of Fig.~\ref{spec}).
\item Looking closer, there is slightly lower variability in the new observation, especially at the highest frequencies (Figs.~\ref{covariance} and \ref{coherence}).
\item The coherence of the high-frequency variability is low for the new observations (and not the old), and therefore we cannot see a clear soft lag at high frequencies in the new observations (Fig.~\ref{coherence} and \ref{high_lagfreq})
\item Both the new and old observations show the typical hard, positive lag at low frequencies, but at the lowest frequencies, there is a break, below which the lag disappears or switches to a negative lag (Figs.~\ref{low_lagfreq} and \ref{low_lagen}).
\item The correlated variability at the Fe~K line is not strong in either observation (Figs.~\ref{covariance} and \ref{cov_rat}), and there is no clear Fe~K lag (Fig.~\ref{high_lagen}).
\end{enumerate}

We discuss the origin of these results in the following subsections.

\subsection{Where is the Fe~K lag?}

The statistical detectability of the lag is determined by the source flux, variability, and observation length.  Given the count rate, excess variance and exposure for MCG-6-30-15, we would expect to find an Fe~K lag, based on detections in 10 other sources thus far. It is not immediately obvious why we do not find an iron K lag in this source.  
%However, what is important to note that while this source has sufficient variability, the lags will only be measured between lightcurves with correlated variability.
The answer may be because there are two different variability mechanisms at play, which have a similar effect on the rms variability, but have different effects on the lag.
In order to measure a lag, the direct emission needs to vary intrinsically, and the reprocessed emission needs to show a linear transformation of that same intrinsic variability.  However, variability can also arise from geometric changes in the position and extent of the direct continuum source.  This will not cause correlated variability in the reprocessed emission, and therefore does not cause a lag measurement. As these two different variability mechanisms have a distinct effect on the measured lag, it is important for us to disentangle the variability, and see how much of it is correlated between the powerlaw and reflection components.

It is generally accepted that the corona varies intrinsically on a variety of timescales. 
% In order to measure time lags between the corona and reprocessed emission, the variability in the corona must be intrisic to the source, and that variability pattern is then also observed in the reprocessed emission.  
This is likely the dominant source of variability in most Seyferts where we have measured Fe~K lags.  In MCG-6-30-15, in addition to the intrinsic, correlated coronal variability, we observe uncorrelated variability between the powerlaw and the reflection.  This was first shown using the 2001 {\em XMM-Newton} data, in which a spectral analysis was performed on several 10-20~ks segments, revealing that the powerlaw flux varied at these timescales, but the reflection did not \citep{fabian03,vaughan04}.  The same result has been showed recently through Principal Component Analysis of the old and new observations \citep{parker14}.

This is attributed to another variability mechanism, namely geometrical changes in the height and extent of the corona.  Observations show that the powerlaw emission is anisotropic, which is a natural result of the corona being very close to the black hole, where gravitational light bending effects are inevitable \citep{miniutti03,miniutti04}.  In the light bending scenario, the powerlaw emission varies because of geometrical changes in the X-ray source, and as the source gets close to the black hole, gravitational light bending will preferentially direct the powerlaw emission in to the black hole and towards the innermost radii of the accretion disc.  This will cause the powerlaw flux to decrease, while the reflected flux varies much less.

%One possible reason for the lack of Fe~K lag in this source is to do with its small black hole mass \citep[$M=2 \times 10^{6} M_{\odot}$][]{zhou10}. MCG-6-30-15 has a lower black hole mass than many other sources with known Fe~K lags.  The lower the mass, the higher frequency we need to probe in order to see the reverberation lag.  In this source, we are closer to the frequency where Poisson noise begins to dominate.  Furthermore the amplitude of the reverberation lag will be smaller in this source than in higher mass sources.  It is possible that the hard lag dominates over the small amplitude soft lag, and therefore we cannot disentange the reverberation in this source. In other, more massive Seyferts, the amplitude of the reverberation lag dominates at high frequencies, and so it is easier to pick out the Fe~K signature.  1H0707-495 is another NLS1 with known Fe~K lag that has a similarly small black hole mass (also $2 \times 10^{6} \mathrm{M}_{\odot}$), however it has an iron abundance of $\sim 10$ times the solar abundance, which perhaps explains why we find the Fe~K lag in 1.3~Ms of data.

These previous spectral timing results showed that reflection is not varying greatly with the powerlaw on long timescales, 10~ks and greater. This corresponds to our very lowest frequencies, but what about at high frequencies, where we expect to see the Fe~K reverberation lag?  The covariance spectrum, which picks out the energy spectrum of just the correlated variability, can help us understand which variability mechanism dominates at a particular energy (Fig.~\ref{covariance}).  The mean spectrum (in black) shows a clear broad Fe~K line \citep{tanaka95,fabian03,marinucci14}.  However, when we look at the spectra of the correlated variability at a number of frequencies, we see the flux peaks at $\sim 2$~keV, and decreases at the energy of the broad Fe~K line, indicating that not all the emission contributing to the broad Fe~K line is correlated with the variability of the corona.  This is the case for all frequencies, even at the highest frequencies, where we would expect to find an Fe~K lag.  This point is made clearer in Fig.~\ref{cov_rat}, where we plot the ratio of the mean spectrum to the covariance spectrum.  This essentially picks out the component that does {\em not} contribute to the correlated variability (and therefore not contributing to the lag). At all frequencies, the shape resembles the broad iron line and soft excess.  The covariance spectrum and covariance ratio are independent ways of showing that on at all timescales probed there is a contribution from reflection that is not correlated with the variable powerlaw.  Similar results from the low-frequency Fourier-resolved spectra of the total variability (i.e. not just correlated variability) have been found by \citet{papadakis05} for the old observations of MCG-6-30-15.  

Our high-frequency covariance spectrum results can be explained by light bending in which the variability on the order of 1~ks and greater is due to geometrical changes in the corona that cause the powerlaw flux to vary more than the reflected flux.  Because there is not much contribution from reflection in the high-frequency emission, we do not see a strong Fe~K lag. The covariance and the lag both show that there is some correlated variability between the powerlaw and the reflection (i.e. that some of the variability is intrinsic to the corona), which is why we measure some soft lag. It is just not enough to produce the strong Fe~K lag feature.    

Another complicating issue here is that MCG-6-30-15 is a lower mass object than many of the other sources with Fe~K lags, and therefore, the amplitude of the reverberation lag will be smaller. Small deviations in the lag are more challenging to detect given the current data quality.

\subsection{What is happening at the lowest frequencies?}

The low frequency behavior of MCG-6-30-15 is quite unlike what is usually seen in variable `bare' Seyferts.  Typically at low frequencies, we see a hard lag that is interpreted phenomonologically as fluctuations in the outer accretion disc that get propagated inwards with the flow. Those fluctuations in the seed photons transfer up to the corona, causing the soft X-ray photons (from larger radii) to respond before hard X-ray photons (produced at smaller radii).  This model, proposed by \citet{kotov01} and applied to AGN by \citet{arevalo06}, is the most common interpretation, but the phenomenon is still not well understood.  In MCG-6-30-15, we find evidence for a hard propagation lag, but that hard lag does not continue to the lowest frequencies.  There is a turnover in the lag that occurs at around 1--2 $\times 10^{-4}$~Hz for both the new and old observation, though the break frequency is slightly lower for the newer observation.  The turnover at low frequencies is present in each individual observation, and therefore does not appear to be due to a transient phenomenon, such as variable neutral absorption from eclipsing clouds along the line of sight.

It is possible that because MCG-6-30-15 is a low mass AGN \citep[$2 \times 10^{6} M_{\odot}$][]{zhou10}, we are able to probe a new regime in the low-frequency lags: the lower limit of the propagation lag.  Variability timescale are shorter for smaller black hole masses, so it is possible that in this source, we are probing frequencies lower than the propagation lags.  In this case, the soft lag we observe could have the same origin as the soft lag at high frequencies.  General Relativisitic ray tracing simulations predict that reverberation will occur between the corona and the accretion disc out to large radii, i.e. tens to hundreds of gravitational radii \citep{reynolds99,cackett14}, so in the absence of a propagation lag, we would observe the reverberation lag down to low frequencies.
% While the disc emissivity is much lower at large radii than within $\sim 10 r_{\mathrm{g}}$, we still do expect there to be a reverberation signal, which would occur at very low frequencies and have a larger amplitude than the soft lag found at high frequencies.  If, in MCG-6-30-15, the propagation lags drop off at the turnover frequency, we may be seeing a signature of reverberation in the lowest frequency range. 
 Clear evidence for this interpretation would be a disc reflection signature in the very-low-frequency lag-energy spectrum, however this can not be determined given the current data quality (even when combining all 600~ks of observations).  Future observational and theoretical work are required in understanding the propagation lags, but what is evident from all observations of X-ray time lags is that the hard lags behave very differently than the soft lags, which show clear indications of relativistic reflection off the inner accretion disc.

\subsection{What has changed between the observations?}

The old observation looks similar to the new observation at low frequencies. Both show a featureless, log-linear lag increasing with energy at frequencies $\sim [2-5] \times 10^{-4}$~Hz (right panel of Fig.~\ref{low_lagen}), and both show the hard lag disappear or soften below $\sim 2 \times 10^{-4}$~Hz (Fig.~\ref{low_lagfreq}).  The low-frequency soft lags, and the frequency of the break are slightly different between old and new observations, but generally, the observations look similar in low-frequency lag, coherence and covariance.  At high frequencies, however, we see that the coherence is lower in the new observation and the significance of the soft lag is much lower.  In addition to this, the mean spectrum appears to be different between the two observations. The right panel of Fig.~\ref{spec} shows that the red wing of the broad Fe~K component is stronger in the old observation than in the new.
This narrower Fe~K line is observed in each orbit of the new observation, regardless of flux.

We can explain these difference as due to changes in the physical extent of the corona.  If the corona is less extended in the old observation, then the reflection fraction will be greater, and the spectrum will show a stronger broad line, especially at the red wing.  Also, for a less extended source the reverberation signature will be more coherent because the reprocessing happens at a smaller region. Whereas, if in the new observation the corona is extended, the signal is smeared out more because the reprocessing is happening from a larger range of radii, and therefore the reprocessed signal is averaged over many different light paths.  The changing lags in MCG-6-30-15 may simply be indicating that the extent of the corona is important and can change significantly.  This extended corona could contribute to smearing out the Fe~K lag. Also, if we are observing a drop-off of the propagation lag at low-frequencies, then it is consistent that the propagation lag would drop off at lower frequencies for the more extended corona (as in Fig.~\ref{low_lagfreq}).  

These results on MCG-6-30-15 suggest that future work modelling the time lags from an extended corona is necessary. Brenneman et al., {\em in prep.}, will present detailed modelling of the iron line profile with a lamppost geometry in order to quantify the variations in coronal height.

\section{Conclusions}
\label{conclusion}

%In this paper we have presented the X-ray time lags from 600~ks of observations of MCG-6-30-15. While the source shows a clear broad Fe~K emission line, there is no strong evidence for a corresponding Fe~K lag in any of the observations. This and the fact that the high-frequency lags change dramatically between the old and new observations may be suggesting that the X-ray source is extended, and that the change in physical extent affects the observed lag. It is also likely that strong light bending effects are important in this source, and will damp out the reflection variability at all frequencies.  

MCG-6-30-15 has given us many insights into the X-ray emitting region around supermassive black holes, and yet there is much still to understand about this intriguing source. The spectral timing analysis here demonstrates some of these complexities.  We have shown that the Fe~K lag is not present in all variable bare Seyferts, which indicates that the importance of understanding the variability mechanisms.  In MCG-6-30-15, geometrical changes in the position of the corona as well as intrinsic changes in the corona itself cause its complex variability structure.  X-ray timing analyses like the one presented here can help decouple these mechanisms.  We also find the first clear evidence for a cut-off in the hard lags, which can help put constraints on our physical models for the propagation lag.  Lastly, the changing lag between 2001 and 2013 suggests that the corona is extended, complicating the observed reverberation signal.  The results presented here illustrate that we need a more sophisticated picture, including an extended X-ray emitting region. 

\section*{Acknowledgements}

We thank an anonymous reviewer for helpful comments. This work is based on observations obtained with {\em XMM-Newton}, an ESA science mission with instruments and contributions directly funded by ESA Member States and NASA.  EK thanks the Gates Cambridge Scholarship. ACF thanks the Royal Society. EK, ACF, AM, GM and WA acknowledge support from the European Union Seventh Framework Programme (FP7/2007-2013) under grant agreement n.312789, StrongGravity.

%\bsp

\label{lastpage}

\end{document}